\documentclass[prl,twocolumn,showpacs,amsmath,amssymb,superscriptaddress]{revtex4-1}
\usepackage{epsfig}
\usepackage{graphicx}
\usepackage{dcolumn}
\usepackage{amsmath}
\usepackage{amssymb}
\usepackage {color}
\begin{document}

\setlength{\parskip}{0cm}
\setlength{\parindent}{0.5em}

 
\title{Restoring Interlayer Josephson Coupling in La$_{1.885}$Ba$_{0.115}$CuO$_4$ by 
Charge Transfer Melting of Stripe Order}

\author{V. Khanna}
\affiliation{Diamond Light Source, Chilton, Didcot, OX11 0DE, United Kingdom}
\affiliation{Max Planck Institute for the Structure and Dynamics of Matter, 22761 Hamburg, Germany}
\affiliation{Department of Physics, Clarendon Laboratory, University of Oxford, Oxford, OX1 3PU, United Kingdom}
\author{R. Mankowsky}
\affiliation{Max Planck Institute for the Structure and Dynamics of Matter, 22761 Hamburg, Germany}
\affiliation{Centre for Free Electron Laser Science, 22761 Hamburg, Germany}
\author{M. Petrich}
\affiliation{Max Planck Institute for the Structure and Dynamics of Matter, 22761 Hamburg, Germany}
\affiliation{Centre for Free Electron Laser Science, 22761 Hamburg, Germany}
\author{H. Bromberger}
\affiliation{Max Planck Institute for the Structure and Dynamics of Matter, 22761 Hamburg, Germany}
\affiliation{Centre for Free Electron Laser Science, 22761 Hamburg, Germany}
\author{S. A. Cavill}
\affiliation{Diamond Light Source, Chilton, Didcot, OX11 0DE, United Kingdom}
\affiliation{Department of Physics, University of York, Heslington, York, YO10 5DD, United Kingdom}
\author{E. M\"ohr-Vorobeva}
\affiliation{Department of Physics, Clarendon Laboratory, University of Oxford, Oxford, OX1 3PU, United Kingdom}
\author{\\D. Nicoletti}
\affiliation{Max Planck Institute for the Structure and Dynamics of Matter, 22761 Hamburg, Germany}
\affiliation{Centre for Free Electron Laser Science, 22761 Hamburg, Germany}
\author{Y. Laplace}
\affiliation{Max Planck Institute for the Structure and Dynamics of Matter, 22761 Hamburg, Germany}
\affiliation{Centre for Free Electron Laser Science, 22761 Hamburg, Germany}
\author{G. D. Gu}
\affiliation{Condensed Matter Physics and Materials Science Department, Brookhaven National Laboratory, Upton, NY, USA}
\author{J. P. Hill}
\affiliation{NSLS-II and Condensed Matter Physics and Materials Science Department, Brookhaven National Laboratory, Upton, New York, USA}
\author{M. F\"orst}
\affiliation{Max Planck Institute for the Structure and Dynamics of Matter, 22761 Hamburg, Germany}
\affiliation{Centre for Free Electron Laser Science, 22761 Hamburg, Germany}
\author{A. Cavalleri}
\altaffiliation[Electronic mail:]{andrea.cavalleri@mpsd.mpg.de}
\affiliation{Max Planck Institute for the Structure and Dynamics of Matter, 22761 Hamburg, Germany}
\affiliation{Department of Physics, Clarendon Laboratory, University of Oxford, Oxford, OX1 3PU, United Kingdom}
\affiliation{Centre for Free Electron Laser Science, 22761 Hamburg, Germany}
\author{S. S. Dhesi}
\altaffiliation[Electronic mail:]{dhesi@diamond.co.uk}
\affiliation{Diamond Light Source, Chilton, Didcot, OX11 0DE, United Kingdom}
\date{\today}

\begin{abstract}

We show that disruption of charge-density-wave (stripe) order by charge transfer excitation, enhances the superconducting phase rigidity in La$_{1.885}$Ba$_{0.115}$CuO$_4$ (LBCO). 
Time-Resolved Resonant Soft X-Ray Diffraction demonstrates that charge order melting
is prompt following near-infrared photoexcitation whereas the crystal structure remains intact for moderate fluences.
THz time-domain spectroscopy reveals that, for the first 2ps following photoexcitation, a
new Josephson Plasma Resonance edge, at higher frequency with respect to the equilibrium edge, is induced
indicating enhanced superconducting interlayer coupling. The fluence dependence
of the charge-order melting and the enhanced superconducting interlayer 
coupling are correlated with a saturation limit of $\sim$0.5mJ/cm$^2$.
Using a combination of x-ray and optical spectroscopies we establish a hierarchy 
of timescales between enhanced superconductivity, melting of charge order and 
rearrangement of the crystal structure.

\end{abstract}

\maketitle

The interplay between superconductivity and 
broken electronic symmetries has emerged as a central theme in cuprate 
physics with increasing reports of charge order in several 
materials \cite{tra1,wu1,ghi1,cha1,com1,sil1,tab1,com2,hay1,fuj1}.
This body of work has lead to a 
growing realization that understanding competing orders may 
be key to developing high-$T_C$ superconductivity further.
An early example of electronic order 
competing with superconductivity is that of charge and
spin (stripe) ordering \cite{tra1,huc1} in underdoped La$_{2-x}$Ba$_{x}$CuO$_4$, where
holes doped into the CuO$_2$ planes order along domain walls, 
separating regions of antiphased antiferromagnetic spin ordering,
below $T\simeq$ 42K. Concomitantly, the 
crystal structure distorts into a low-temperature 
tetragonal (LTT) phase which is assumed to align the
hole-rich domain walls along the Cu-O-Cu bond direction
with a doping dependent stripe modulation \cite{fin1,cra1,suz1}.
The emergence of superconductivity from this stripe phase for 
0.09 $\lesssim$ x $\lesssim$ 0.16 follows a peculiar 
double-dome phase boundary \cite{moo1}, with 
the superconducting transition temperature, T$_C$, greatly 
suppressed for $x=$1/8. 

A first step in disentangling the
hierachy of cause and effect demonstrated
that stripe ordering persists in the 
absence of the LTT distortion \cite{huc2,tha1}. In a similar fashion
mid-IR light has also been used to induce lattice distortions
in La$_{1.875}$Ba$_{0.125}$CuO$_4$ revealing that
the LTT distortion and stripe ordering evolve as distinct non-equilibrium
phases with different timescales \cite{for1}. More recent work, argues that
pair-density waves in the CuO$_2$ planes  completely suppress Josephson coupling between neighboring 
planes \cite{li1,ber1}. The loss of interlayer coupling is, however,
predicted to be highly sensitive to topological defects
resulting in a complex phase diagram 
with the emergence of new and novel superconducting
phases with increasing defect density \cite{he1}.

Here, we reveal how puncturing stripe order, with charge transfer defects introduced by near-infrared optical pulses, 
can dynamically enhance the superconducting order at the expense 
of stripe order. Specifically, we combine photoexcitation, Time-Resolved 
Resonant Soft X-ray Diffraction (TR-RSXD) and THz time-domain spectroscopy to demonstrate that
ultrafast disruption of the stripe ordered phase promptly 
enhances interlayer Josephson coupling in La$_{1.885}$Ba$_{0.115}$CuO$_4$ (LBCO).
Furthermore, we show that the fluence dependence of the 
enhanced interlayer Josephson coupling follows closely that of 
the stripe order melting and not that of lattice 
rearrangement.

We studied LBCO, for which bulk superconductivity 
develops at T$_C$ $\simeq$ 13K, spin-ordering 
at T$_{SO}$ $\simeq$ 42K and charge ordering 
along with the LTT distortion 
at T$_{CO}$ $\simeq$ T$_{LTT}$ $\simeq$ 53K \cite{huc1}.
Stripe ordering in LBCO
is described using a charge-density wave with incommensurate 
wave vector (0.23 0 0.65) \cite{abb1}. 
The LTT structural distortion involves a
buckling of the Cu-O-Cu bonds in the CuO$_2$ planes
and a tilting of the CuO$_6$ octahedra which allows
the (0 0 1) reflection \cite{fin1}. The intensities of the 
(0.23 0 0.65) and (0 0 1) diffraction peaks are therefore
direct probes of the degree of stripe ordering and LTT distortion.

Single crystals of LBCO were 
grown using the traveling-solvent-floating 
zone method \cite{huc1}. Two samples from the same batch were used; 
one was cleaved along the $ab$ surface for TR-RSXD, the other cut and polished 
to give an $ac$ surface for THz spectroscopy. All measurements 
were carried out in the superconducting state (T$\simeq$5K).
The 1.55eV (800nm) pump pulses used to photoexcite the sample were 
$\sigma$-polarized, parallel to the Cu-O-Cu bond direction. The 
penetration depth at 1.55 eV and $\sim$530 eV, estimated from LBCO 
equilibrium optical properties, is similar 
($\sim$200 nm) \cite{hom1,han1}, so that a homogeneous photoexcited 
volume was probed. The 1.55eV pump 
pulses from a Ti:Sa amplifier running at 22 KHz were 
focused down to a spot size of $\sim$200$\mu$m  (FWHM) onto 
the sample, whilst the X-ray spot size was $\sim$100$\mu$m.
The diffraction peak intensities were measured 
using a gated micro-channel plate, which was insensitive to 
the 1.55eV pump.

Figure 1 shows the energy dependence of the 
(0.23 0 0.65) stripe and (0 0 1) structural diffraction peaks 
as well as the x-ray absorption spectroscopy (XAS) spectrum across the O $K$-edge 
recorded at T=5K. The resonances at 528.6 eV, 
along with the weak shoulder at 530.2 eV,
correspond to transitions into the
states associated with the doped holes and the
upper Hubbard band \cite{fin2}. The (0 0 1) diffraction peak has a strong 
resonance at the O $K$-edge, centered at 532.4 eV 
corresponding to resonant transitions into La-O hybridized 
states \cite{fin1}.

\begin{figure}[!t]
\includegraphics[width=\columnwidth]{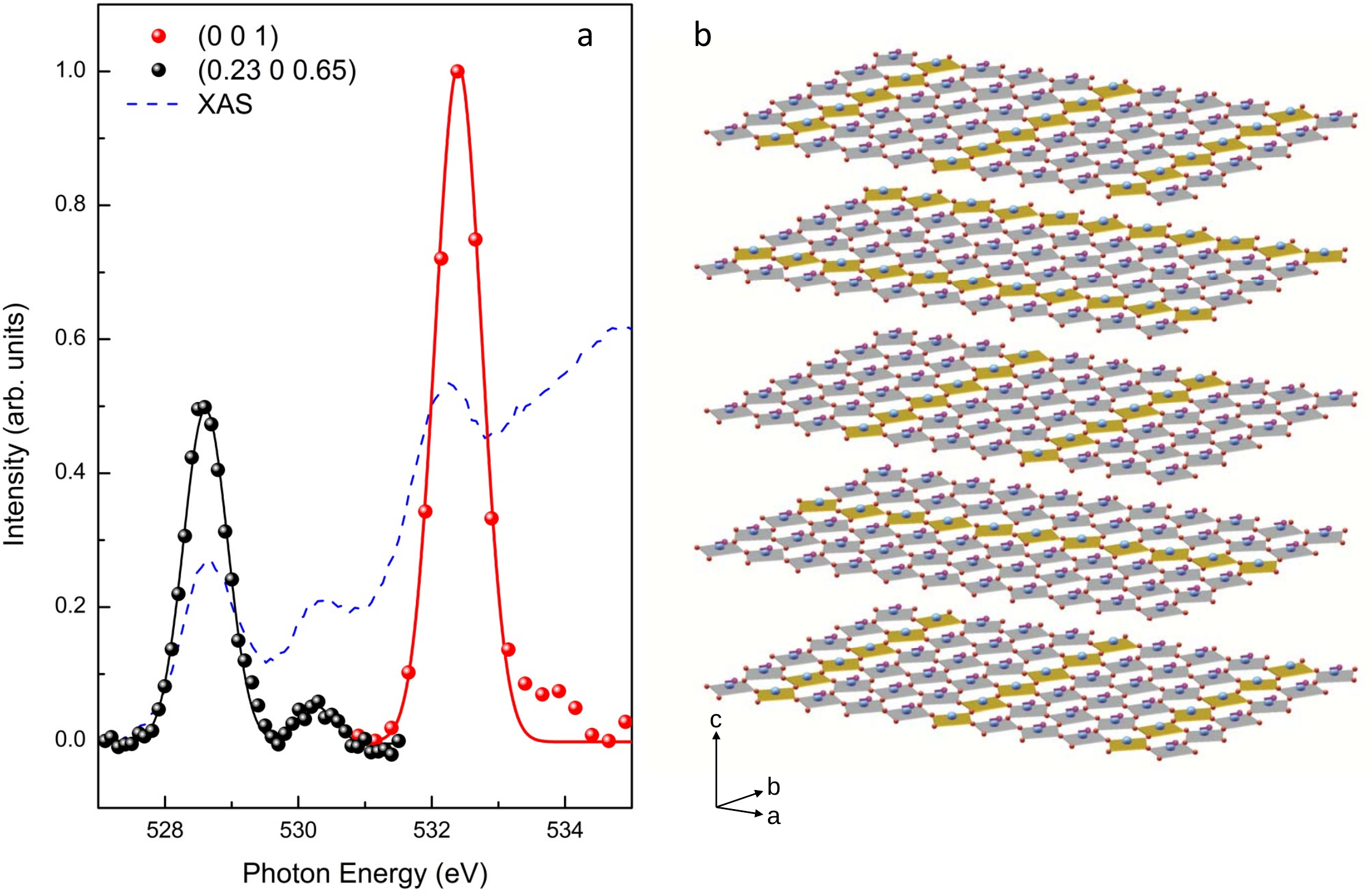}
\caption{ (Color online) (a) XAS spectrum for LBCO over the O K-edge 
(dashed blue line) along with the (0.23 0 0.65) 
charge-ordering diffraction peak (black circles) 
and (001) LTT distortion diffraction peak (red circles). 
The lines represent Lorentzian fits to the data. 
The (0.23 0 0.65) peak has been scaled up by a 
factor of 3000. (b) Schematic of the 
stripe ordering (yellow polygons) in the CuO$_2$ planes.}
\label{fig:fig1}
\end{figure}

We first discuss the TR-RSXD results.
The temporal evolution of both the stripe and LTT phase
after photoexcitation was investigated using ultrafast 
changes in the (0.23 0 0.65) and (0 0 1) diffraction peaks
relative to the fluorescence background. 
Figure 2 shows the change in diffraction peak intensities ($\Delta$I$_{\tau}$/I$_0$) as a 
function of time delay ($\tau$), for an excitation fluence of 1.6 mJ/cm$^2$. 
The results were recorded with the storage ring 
operating in low-$\alpha$ mode, for which the 
longitudinal width of the electron bunch was compressed 
to give a temporal resolution of $\sim$7ps at the expense of photon flux. 

\begin{figure}[!t]
\includegraphics[width=.45\textwidth]{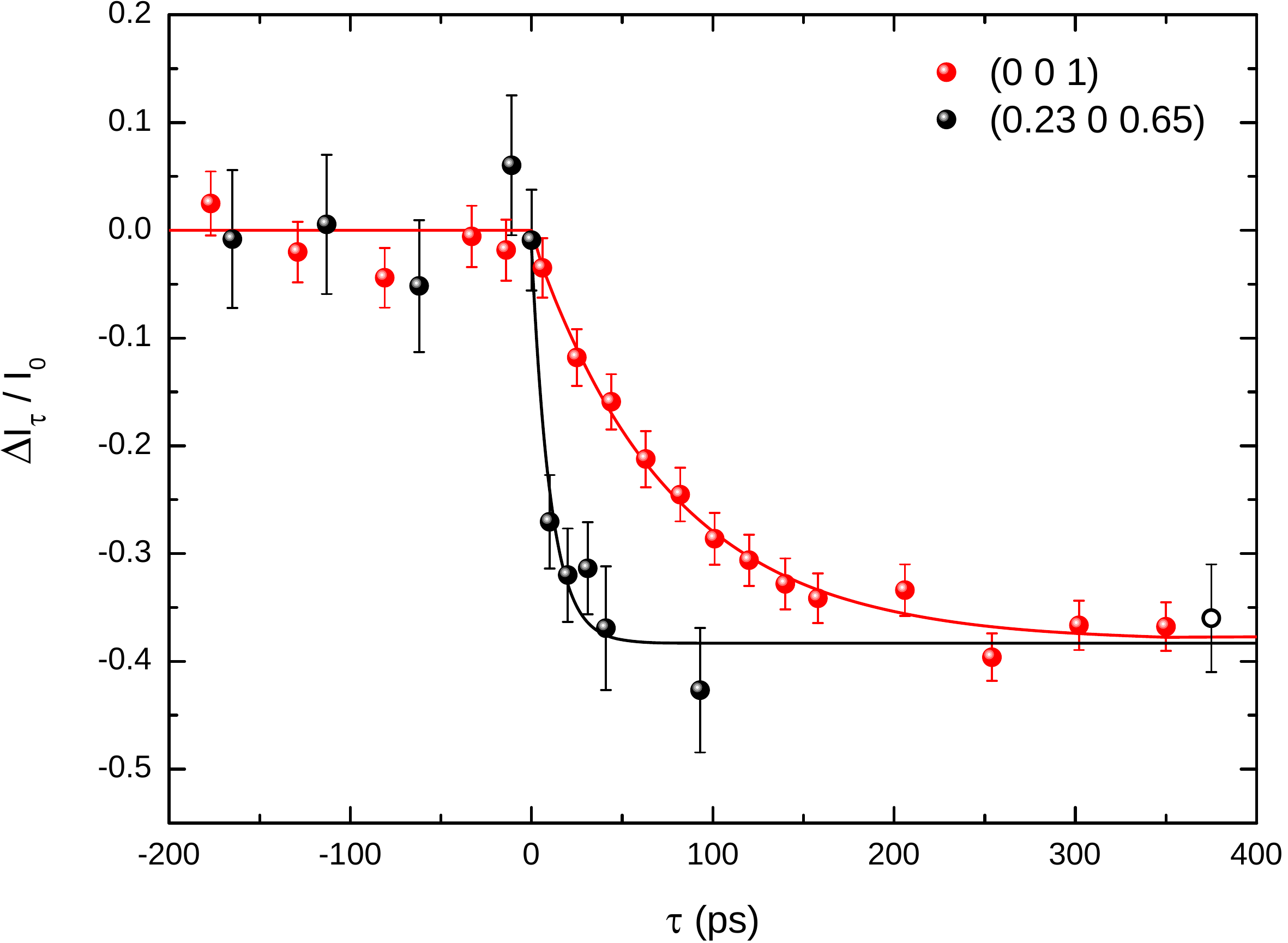}
\caption{ (Color online) Intensity changes in the (0.23 0 0.65) 
charge-ordering diffraction peak (solid black circles) and (001) 
LTT distortion  diffraction peak (solid red circles) following near-infrared 
photoexcitation using a pump fluence of 1.6 mJ/cm$^2$. 
The data were measured in low-$\alpha$ mode with an 
X-ray pulse width of $\sim$7ps (FWHM) except
for the open circle which represents the change in intensity of the
(0.23 0 0.65) diffraction peak for $\tau$=375ps recorded in hybrid mode
with an X-ray pulse width of $\sim$60ps (FWHM). The 
solid lines represent fits to the data using an exponential function.}
\end{figure}

For the same excitation 
fluence, both the (0.23 0 0.65) and (0 0 1) peaks are 
reduced in intensity by $\sim$40\%. However, the temporal 
response after photoexcitation is different. A fit to the stripe
peak data (Fig. 2, black line) gives a time constant of 10$\pm$3 ps and
is limited by the X-ray pulse width in the low-$\alpha$ mode of operation. 
However, the decay in the stripe peak is likely to be prompt, occurring within only a 
few hundred femtoseconds of photoexcitation \cite{for1}. The (0 0 1) peak, 
on the other hand, is observed to decrease over a much slower 
timescale with an exponential fit (Fig. 2, red line) yielding a time constant of 
77$\pm$7 ps. We note here that the different responses of the two peaks precludes
loss of intensity from sample heating from the laser pulses since T$_{CO}$ $\simeq$T$_{LTT}$.
Previous studies using mid-infrared excitation of in-plane stretching 
modes have shown enhancement of interlayer coupling 
in a number of cuprates \cite{fau1,hu1}, but 
little is known regarding the fate of the charge ordered 
phase in the non-equilibrium state \cite{for1,for2}. 
Figure 2, demonstrates that photoexitation 
using near-infrared 1.55eV pulses creates a charge transfer non-equilibrium 
phase in which the LTT distortion remains intact, but 
the stripe ordering is strongly suppressed. This then gives a unique 
system with which to explore the emerging dynamics of superconductivity 
once stripe order is disrupted.

The photoinduced dynamics of the superconducting 
condensate were measured using transient reflectivity 
at THz frequencies. The sample was 
excited under the same conditions used 
in the TR-RSXD work with near-infrared 1.55eV laser pulses.
The weak interlayer superconducting coupling of LBCO results in an equilibrium 
Josephson Plasmon Resonance (JPR) at $\sim$0.2THz \cite{equJPR,nic1} in the
$c$-axis reflectivity and is shown as the gray line in Fig. 3(a). Superconducting transport
is also observed as a small low-frequency divergence of the imaginary
part of the conductivity ($\sigma_2$) and is shown as the grey line in Fig. 3(b).

\begin{figure}[!t]
\includegraphics[width=\columnwidth]{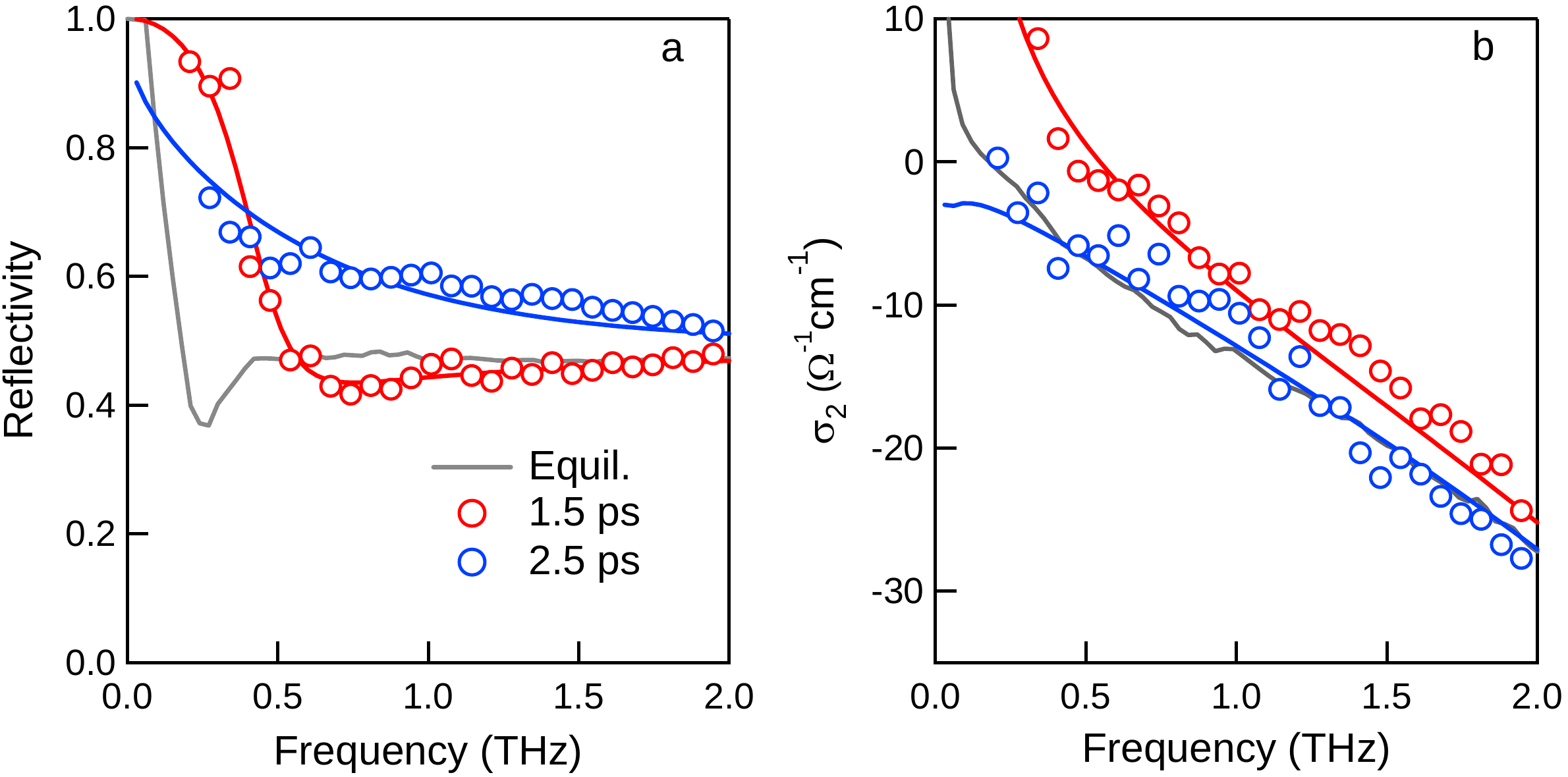}
\caption{ (Color online) 
(a) Normal-incidence reflectivity and (b) imaginary conductivity ($\sigma_2$) from LBCO measured with THz
time-domain spectroscopy 1.5 ps (red circles) and 2.5 ps (blue circles) after near-infrared optical excitation.
The same quantities measured at equilibrium are displayed as gray lines. Red and blue lines are fits
to the data performed with a superconducting and Drude model, respectively.  } 
\end{figure}

The transient optical properties of the photoexcited 
LBCO were measured in the 0.15-2 THz spectral range for different pump 
fluences as a function of $\tau$, with a time 
resolution of $\sim$350fs, determined by the inverse bandwidth 
of the THz probe pulse. The mismatch between the penetration 
depth of the near-infrared 1.55eV pump ($\sim$0.1$\mu$m) and that of 
the THz probe ($\sim$50-500$\mu $m) was taken into account using a 
multilayer model \cite{hu1,mismatch1}. The $c$-axis transient reflectivity 
and $\sigma_2$ spectra, measured 1.5ps after photoexcitation, 
are shown in Fig. 3 (red circles) for a pump fluence of 1.6 mJ/cm$^2$. The 
JPR displays a prompt blue-shift from $\sim$0.25 THz to $\sim$0.5 THz (Fig. 3(a), red circles). 
Correspondingly, an enhancement of $\sigma_2$
is indicative of an increase in interlayer 
Josephson coupling (Fig. 3(b), red circles). At $\tau$=2.5ps  we
observe a relaxation to a state in which coherence 
is reduced, characterized by a broader edge in 
reflectivity (Fig. 3(a), blue circles) and the absence of a divergence in $\sigma_2$ (Fig. 3(b), blue circles).
In analogy with previous optical results for LBCO after photoexcitation perpendicular 
to the CuO$_2$ planes \cite{nic1,cas1}, the transient spectra could be fitted
assuming a superconducting model for $\tau$=1.5ps (Fig. 3, red lines), 
while in the relaxed state a Drude model with 
a finite carrier scattering time in the picosecond 
range had to be employed (Fig. 3, blue lines). Notably, in the case of in-plane pumping 
explored here, the blue-shift of the JPR is less marked and the 
relaxation to the incoherent state is faster with respect to the case of $c$-axis
pumping \cite{nic1}, most likely due 
to stronger coupling to quasiparticle excitation in the planes.
The enhancement of the superconducting phase using
charge transfer excitation is surprising given that the process generates high energy electrons that relax
via the emission of phonons which subsequently destroy the Cooper
pairs \cite{kus1,sma1,cor1}. For LBCO the disruption of interlayer superconductivity 
due to pair-density waves \cite{ber1} is presumably removed following charge transfer leading to
an enhancement of the superconducting phase before its destruction
from the creation of quasiparticles.

\begin{figure}
\includegraphics[width=.4\textwidth]{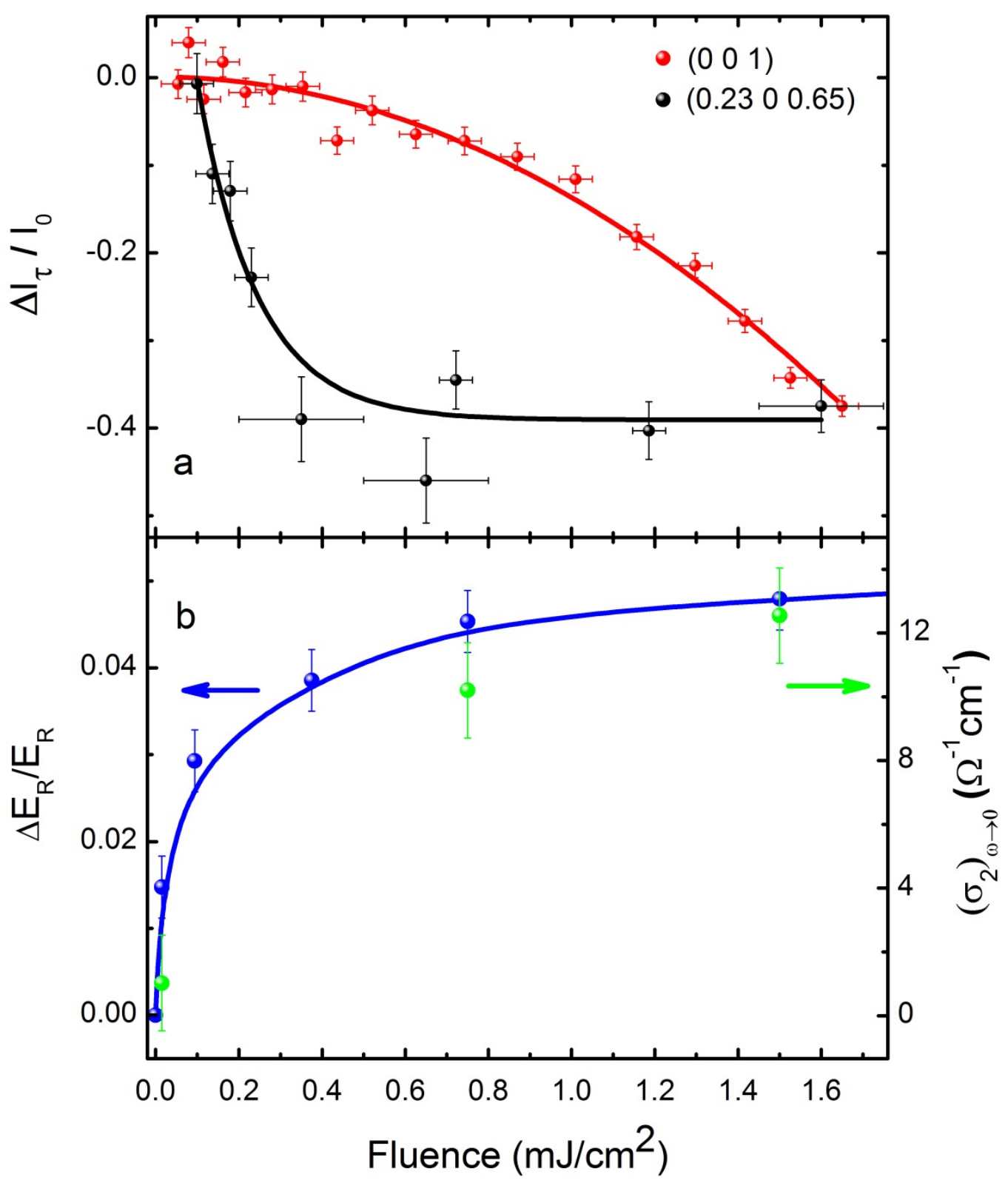}
\caption{ (Color online) (a) Fluence dependence of the (0.23 0 0.65) 
charge-ordering diffraction peak (solid black circles) and (001) LTT distortion 
diffraction peak (solid red circles) intensity after photoexcitation 
with $\tau$=350ps. The error in the fluence indicates 
the uncertainty in the pump beam diameter at the sample position.
The red line shows a quadratic fit whereas
the black line shows an exponential fit to the data.
(b) Changes in the electric field, $\Delta$E$_R$/E$_R$ (solid blue circles) 
and ($\sigma_2$)$_{\omega\rightarrow 0}$ (solid green circles) as a function of fluence 
(Frequency = 0.2THz, $\tau$ = 1.5ps). The blue line shows a fit using an exponential function.}
\end{figure}     

The measurement of both TR-RSXD and THz time-domain reflectivity, under 
the same near-infrared excitation conditions, allows the first direct 
comparison of the time scales involved in the 
dynamics. The interlayer coupling enhancement, which develops 
in $\sim$1ps following photoexcitation, is strongly connected with the
stripe order melting shown in Fig. 2. On the other hand, the lattice dynamics
start to develop on considerably longer timescales ($>$10ps).

Further insights regarding the enhanced interlayer coupling
and the disruption of stripe order can be retrieved 
from the pump fluence dependence of the intensities measure using
TR-RSXD and THz spectroscopy. Figure 4(a) shows $\Delta$I$_{\tau}$/I$_0$ 
for the stripe and LTT distortion diffraction peaks 
as a function of fluence for $\tau$=350ps. 
The results were recorded with the storage ring operating
in hybrid mode with a temporal resolution of $\sim$60ps. This establishes
that at $\sim$0.5mJ/cm$^2$ the melting of the stripe 
ordering saturates whereas the LTT distortion remains 
largely intact which should remain the case for the shortest
picosecond timescales inaccessible for synchrotron based TR-RSXD.
These changes can then be compared with the 
fluence-dependent Josephson interlayer coupling 
enhancement (Fig. 4b), estimated by the increase 
in the low-frequency limit of the imaginary
part of the conductivity ($\sigma_2$)$_{\omega\rightarrow 0}$ 
and by the change in the reflected electric field ($\Delta$E$_R$/E$_R$) 
at the peak of the response ($\tau$=1.5ps). These results
confirm the causal link between stripe melting and 
superconductivity enhancement, as both the (0.23 0 0.65) 
diffraction peak and the $\sigma_2$ response show a clear saturation 
at $\sim$0.5 mJ/cm$^2$ fluence, which is not present for 
the (0 0 1) peak.

The results show that charge transfer destruction of stripe 
order enhances interlayer superconducting coupling 
in La$_{1.885}$Ba$_{0.115}$CuO$_4$. The fluence dependence of the charge transfer
collapse of stripe order along with the concomitant emergence
of an improved superconducting phase should encourage
further work to reveal the nature of the 
charge transfer process \cite{mag1}.
Systematic studies of the conditions required to depress stripe
order using charge transfer may become effective for high-speed 
devices in which the electrical, optical and magnetic 
properties of highly-correlated materials can be modified 
using near-infrared light.

\begin{acknowledgments}
We thank Diamond Light Source for the provision of beamtime
under proposal numbers SI-7497, SI-7942 and SI-8207.
The research leading to these results has received 
funding from the European Research Council under the 
European Union's Seventh Framework 
Programme (FP7/2007-2013) / ERC Grant Agreement n� 319286 (Q-MAC).
Work at Brookhaven was supported by the Office of Basic Energy Sciences (BES), Division of Materials Sciences and Engineering, U.S. Department of Energy (DOE), through Contract No. DE-SC00112704.
\end{acknowledgments}

\bibliographystyle{apsrev4-1}

\bibliography{LBCO_TRSXD_3}

\begin{thebibliography}{36}%
\makeatletter
\providecommand \@ifxundefined [1]{%
 \@ifx{#1\undefined}
}%
\providecommand \@ifnum [1]{%
 \ifnum #1\expandafter \@firstoftwo
 \else \expandafter \@secondoftwo
 \fi
}%
\providecommand \@ifx [1]{%
 \ifx #1\expandafter \@firstoftwo
 \else \expandafter \@secondoftwo
 \fi
}%
\providecommand \natexlab [1]{#1}%
\providecommand \enquote  [1]{``#1''}%
\providecommand \bibnamefont  [1]{#1}%
\providecommand \bibfnamefont [1]{#1}%
\providecommand \citenamefont [1]{#1}%
\providecommand \href@noop [0]{\@secondoftwo}%
\providecommand \href [0]{\begingroup \@sanitize@url \@href}%
\providecommand \@href[1]{\@@startlink{#1}\@@href}%
\providecommand \@@href[1]{\endgroup#1\@@endlink}%
\providecommand \@sanitize@url [0]{\catcode `\\12\catcode `\$12\catcode
  `\&12\catcode `\#12\catcode `\^12\catcode `\_12\catcode `\%12\relax}%
\providecommand \@@startlink[1]{}%
\providecommand \@@endlink[0]{}%
\providecommand \url  [0]{\begingroup\@sanitize@url \@url }%
\providecommand \@url [1]{\endgroup\@href {#1}{\urlprefix }}%
\providecommand \urlprefix  [0]{URL }%
\providecommand \Eprint [0]{\href }%
\providecommand \doibase [0]{http://dx.doi.org/}%
\providecommand \selectlanguage [0]{\@gobble}%
\providecommand \bibinfo  [0]{\@secondoftwo}%
\providecommand \bibfield  [0]{\@secondoftwo}%
\providecommand \translation [1]{[#1]}%
\providecommand \BibitemOpen [0]{}%
\providecommand \bibitemStop [0]{}%
\providecommand \bibitemNoStop [0]{.\EOS\space}%
\providecommand \EOS [0]{\spacefactor3000\relax}%
\providecommand \BibitemShut  [1]{\csname bibitem#1\endcsname}%
\let\auto@bib@innerbib\@empty
\bibitem [{\citenamefont {Tranquada}\ \emph {et~al.}(1995)\citenamefont
  {Tranquada}, \citenamefont {Sternlieb}, \citenamefont {Axe}, \citenamefont
  {Nakamura},\ and\ \citenamefont {Uchida}}]{tra1}%
  \BibitemOpen
  \bibfield  {author} {\bibinfo {author} {\bibfnamefont {J.~M.}\ \bibnamefont
  {Tranquada}}, \bibinfo {author} {\bibfnamefont {B.~J.}\ \bibnamefont
  {Sternlieb}}, \bibinfo {author} {\bibfnamefont {J.~D.}\ \bibnamefont {Axe}},
  \bibinfo {author} {\bibfnamefont {Y.}~\bibnamefont {Nakamura}}, \ and\
  \bibinfo {author} {\bibfnamefont {S.}~\bibnamefont {Uchida}},\ }\href@noop {}
  {\bibfield  {journal} {\bibinfo  {journal} {Nature (London)}\ }\textbf
  {\bibinfo {volume} {375}},\ \bibinfo {pages} {561} (\bibinfo {year}
  {1995})}\BibitemShut {NoStop}%
\bibitem [{\citenamefont {Wu}\ \emph {et~al.}(2011)\citenamefont {Wu},
  \citenamefont {Mayaffre}, \citenamefont {Kr\"amer}, \citenamefont
  {Horvati\'c}, \citenamefont {Berthier}, \citenamefont {Hardy}, \citenamefont
  {Liang}, \citenamefont {Bonn},\ and\ \citenamefont {Julien}}]{wu1}%
  \BibitemOpen
  \bibfield  {author} {\bibinfo {author} {\bibfnamefont {T.}~\bibnamefont
  {Wu}}, \bibinfo {author} {\bibfnamefont {H.}~\bibnamefont {Mayaffre}},
  \bibinfo {author} {\bibfnamefont {S.}~\bibnamefont {Kr\"amer}}, \bibinfo
  {author} {\bibfnamefont {M.}~\bibnamefont {Horvati\'c}}, \bibinfo {author}
  {\bibfnamefont {C.}~\bibnamefont {Berthier}}, \bibinfo {author}
  {\bibfnamefont {W.~N.}\ \bibnamefont {Hardy}}, \bibinfo {author}
  {\bibfnamefont {R.}~\bibnamefont {Liang}}, \bibinfo {author} {\bibfnamefont
  {D.~A.}\ \bibnamefont {Bonn}}, \ and\ \bibinfo {author} {\bibfnamefont
  {M.~H.}\ \bibnamefont {Julien}},\ }\href@noop {} {\bibfield  {journal}
  {\bibinfo  {journal} {Nature (London)}\ }\textbf {\bibinfo {volume} {477}},\
  \bibinfo {pages} {191} (\bibinfo {year} {2011})}\BibitemShut {NoStop}%
\bibitem [{\citenamefont {Ghiringhelli}\ \emph {et~al.}(2012)\citenamefont
  {Ghiringhelli}, \citenamefont {Le~Tacon}, \citenamefont {Minola},
  \citenamefont {Blanco-Canosa}, \citenamefont {Mazzoli}, \citenamefont
  {Brookes}, \citenamefont {De~Luca}, \citenamefont {Frano}, \citenamefont
  {Hawthorn}, \citenamefont {He}, \citenamefont {Loew}, \citenamefont
  {Moretti~Sala}, \citenamefont {Peets}, \citenamefont {Salluzzo},
  \citenamefont {Schierle}, \citenamefont {Sutarto}, \citenamefont {Sawatzky},
  \citenamefont {Weschke}, \citenamefont {Keimer},\ and\ \citenamefont
  {Braicovich}}]{ghi1}%
  \BibitemOpen
  \bibfield  {author} {\bibinfo {author} {\bibfnamefont {G.}~\bibnamefont
  {Ghiringhelli}}, \bibinfo {author} {\bibfnamefont {M.}~\bibnamefont
  {Le~Tacon}}, \bibinfo {author} {\bibfnamefont {M.}~\bibnamefont {Minola}},
  \bibinfo {author} {\bibfnamefont {S.}~\bibnamefont {Blanco-Canosa}}, \bibinfo
  {author} {\bibfnamefont {C.}~\bibnamefont {Mazzoli}}, \bibinfo {author}
  {\bibfnamefont {N.~B.}\ \bibnamefont {Brookes}}, \bibinfo {author}
  {\bibfnamefont {G.~M.}\ \bibnamefont {De~Luca}}, \bibinfo {author}
  {\bibfnamefont {A.}~\bibnamefont {Frano}}, \bibinfo {author} {\bibfnamefont
  {D.~G.}\ \bibnamefont {Hawthorn}}, \bibinfo {author} {\bibfnamefont
  {F.}~\bibnamefont {He}}, \bibinfo {author} {\bibfnamefont {T.}~\bibnamefont
  {Loew}}, \bibinfo {author} {\bibfnamefont {M.}~\bibnamefont {Moretti~Sala}},
  \bibinfo {author} {\bibfnamefont {D.~C.}\ \bibnamefont {Peets}}, \bibinfo
  {author} {\bibfnamefont {M.}~\bibnamefont {Salluzzo}}, \bibinfo {author}
  {\bibfnamefont {E.}~\bibnamefont {Schierle}}, \bibinfo {author}
  {\bibfnamefont {R.}~\bibnamefont {Sutarto}}, \bibinfo {author} {\bibfnamefont
  {G.~A.}\ \bibnamefont {Sawatzky}}, \bibinfo {author} {\bibfnamefont
  {E.}~\bibnamefont {Weschke}}, \bibinfo {author} {\bibfnamefont
  {B.}~\bibnamefont {Keimer}}, \ and\ \bibinfo {author} {\bibfnamefont
  {L.}~\bibnamefont {Braicovich}},\ }\href@noop {} {\bibfield  {journal}
  {\bibinfo  {journal} {Science}\ }\textbf {\bibinfo {volume} {337}},\ \bibinfo
  {pages} {821} (\bibinfo {year} {2012})}\BibitemShut {NoStop}%
\bibitem [{\citenamefont {Chang}\ \emph {et~al.}(2012)\citenamefont {Chang},
  \citenamefont {Blackburn}, \citenamefont {Holmes}, \citenamefont
  {Christensen}, \citenamefont {Larsen}, \citenamefont {Mesot}, \citenamefont
  {Liang}, \citenamefont {Bonn}, \citenamefont {Hardy}, \citenamefont
  {Watenphul}, \citenamefont {Zimmermann}, \citenamefont {Forgan},\ and\
  \citenamefont {Hayden}}]{cha1}%
  \BibitemOpen
  \bibfield  {author} {\bibinfo {author} {\bibfnamefont {J.}~\bibnamefont
  {Chang}}, \bibinfo {author} {\bibfnamefont {E.}~\bibnamefont {Blackburn}},
  \bibinfo {author} {\bibfnamefont {A.~T.}\ \bibnamefont {Holmes}}, \bibinfo
  {author} {\bibfnamefont {N.~B.}\ \bibnamefont {Christensen}}, \bibinfo
  {author} {\bibfnamefont {J.}~\bibnamefont {Larsen}}, \bibinfo {author}
  {\bibfnamefont {J.}~\bibnamefont {Mesot}}, \bibinfo {author} {\bibfnamefont
  {R.}~\bibnamefont {Liang}}, \bibinfo {author} {\bibfnamefont {D.~A.}\
  \bibnamefont {Bonn}}, \bibinfo {author} {\bibfnamefont {W.~N.}\ \bibnamefont
  {Hardy}}, \bibinfo {author} {\bibfnamefont {A.}~\bibnamefont {Watenphul}},
  \bibinfo {author} {\bibfnamefont {M.~v.}\ \bibnamefont {Zimmermann}},
  \bibinfo {author} {\bibfnamefont {E.~M.}\ \bibnamefont {Forgan}}, \ and\
  \bibinfo {author} {\bibfnamefont {S.~M.}\ \bibnamefont {Hayden}},\
  }\href@noop {} {\bibfield  {journal} {\bibinfo  {journal} {Nature Phys.}\
  }\textbf {\bibinfo {volume} {8}},\ \bibinfo {pages} {871} (\bibinfo {year}
  {2012})}\BibitemShut {NoStop}%
\bibitem [{\citenamefont {Comin}\ \emph {et~al.}(2014)\citenamefont {Comin},
  \citenamefont {Frano}, \citenamefont {Yee}, \citenamefont {Yoshida},
  \citenamefont {Eisaki}, \citenamefont {Schierle}, \citenamefont {Weschke},
  \citenamefont {Sutarto}, \citenamefont {He}, \citenamefont {Soumyanarayanan},
  \citenamefont {He}, \citenamefont {Le~Tacon}, \citenamefont {Elfimov},
  \citenamefont {Hoffman}, \citenamefont {Sawatzky}, \citenamefont {Keimer},\
  and\ \citenamefont {Damascelli}}]{com1}%
  \BibitemOpen
  \bibfield  {author} {\bibinfo {author} {\bibfnamefont {R.}~\bibnamefont
  {Comin}}, \bibinfo {author} {\bibfnamefont {A.}~\bibnamefont {Frano}},
  \bibinfo {author} {\bibfnamefont {M.~M.}\ \bibnamefont {Yee}}, \bibinfo
  {author} {\bibfnamefont {Y.}~\bibnamefont {Yoshida}}, \bibinfo {author}
  {\bibfnamefont {H.}~\bibnamefont {Eisaki}}, \bibinfo {author} {\bibfnamefont
  {E.}~\bibnamefont {Schierle}}, \bibinfo {author} {\bibfnamefont
  {E.}~\bibnamefont {Weschke}}, \bibinfo {author} {\bibfnamefont
  {R.}~\bibnamefont {Sutarto}}, \bibinfo {author} {\bibfnamefont
  {F.}~\bibnamefont {He}}, \bibinfo {author} {\bibfnamefont {A.}~\bibnamefont
  {Soumyanarayanan}}, \bibinfo {author} {\bibfnamefont {Y.}~\bibnamefont {He}},
  \bibinfo {author} {\bibfnamefont {M.}~\bibnamefont {Le~Tacon}}, \bibinfo
  {author} {\bibfnamefont {I.~S.}\ \bibnamefont {Elfimov}}, \bibinfo {author}
  {\bibfnamefont {J.~E.}\ \bibnamefont {Hoffman}}, \bibinfo {author}
  {\bibfnamefont {G.~A.}\ \bibnamefont {Sawatzky}}, \bibinfo {author}
  {\bibfnamefont {B.}~\bibnamefont {Keimer}}, \ and\ \bibinfo {author}
  {\bibfnamefont {A.}~\bibnamefont {Damascelli}},\ }\href@noop {} {\bibfield
  {journal} {\bibinfo  {journal} {Science}\ }\textbf {\bibinfo {volume}
  {343}},\ \bibinfo {pages} {390} (\bibinfo {year} {2014})}\BibitemShut
  {NoStop}%
\bibitem [{\citenamefont {da~Silva~Neto}\ \emph {et~al.}(2014)\citenamefont
  {da~Silva~Neto}, \citenamefont {Aynajian}, \citenamefont {Frano},
  \citenamefont {Comin}, \citenamefont {Schierle}, \citenamefont {Weschke},
  \citenamefont {Gyenis}, \citenamefont {Wen}, \citenamefont {Schneeloch},
  \citenamefont {Xu}, \citenamefont {Ono}, \citenamefont {Gu}, \citenamefont
  {Le~Tacon},\ and\ \citenamefont {Yazdani}}]{sil1}%
  \BibitemOpen
  \bibfield  {author} {\bibinfo {author} {\bibfnamefont {E.~H.}\ \bibnamefont
  {da~Silva~Neto}}, \bibinfo {author} {\bibfnamefont {P.}~\bibnamefont
  {Aynajian}}, \bibinfo {author} {\bibfnamefont {A.}~\bibnamefont {Frano}},
  \bibinfo {author} {\bibfnamefont {R.}~\bibnamefont {Comin}}, \bibinfo
  {author} {\bibfnamefont {E.}~\bibnamefont {Schierle}}, \bibinfo {author}
  {\bibfnamefont {E.}~\bibnamefont {Weschke}}, \bibinfo {author} {\bibfnamefont
  {A.}~\bibnamefont {Gyenis}}, \bibinfo {author} {\bibfnamefont
  {J.}~\bibnamefont {Wen}}, \bibinfo {author} {\bibfnamefont {J.}~\bibnamefont
  {Schneeloch}}, \bibinfo {author} {\bibfnamefont {Z.}~\bibnamefont {Xu}},
  \bibinfo {author} {\bibfnamefont {S.}~\bibnamefont {Ono}}, \bibinfo {author}
  {\bibfnamefont {G.}~\bibnamefont {Gu}}, \bibinfo {author} {\bibfnamefont
  {M.}~\bibnamefont {Le~Tacon}}, \ and\ \bibinfo {author} {\bibfnamefont
  {A.}~\bibnamefont {Yazdani}},\ }\href@noop {} {\bibfield  {journal} {\bibinfo
   {journal} {Science}\ }\textbf {\bibinfo {volume} {343}},\ \bibinfo {pages}
  {393} (\bibinfo {year} {2014})}\BibitemShut {NoStop}%
\bibitem [{\citenamefont {Tabis}\ \emph {et~al.}(2014)\citenamefont {Tabis},
  \citenamefont {Li}, \citenamefont {Le~Tacon}, \citenamefont {Braicovich},
  \citenamefont {Kreyssig}, \citenamefont {Minola}, \citenamefont {Dellea},
  \citenamefont {Weschke}, \citenamefont {Veit}, \citenamefont {Ramazanoglu},
  \citenamefont {Goldman}, \citenamefont {Schmitt}, \citenamefont
  {Ghiringhelli}, \citenamefont {Bari\v{s}i\'c}, \citenamefont {Chan},
  \citenamefont {Dorow}, \citenamefont {Yu}, \citenamefont {Zhao},
  \citenamefont {Keimer},\ and\ \citenamefont {Greven}}]{tab1}%
  \BibitemOpen
  \bibfield  {author} {\bibinfo {author} {\bibfnamefont {W.}~\bibnamefont
  {Tabis}}, \bibinfo {author} {\bibfnamefont {Y.}~\bibnamefont {Li}}, \bibinfo
  {author} {\bibfnamefont {M.}~\bibnamefont {Le~Tacon}}, \bibinfo {author}
  {\bibfnamefont {L.}~\bibnamefont {Braicovich}}, \bibinfo {author}
  {\bibfnamefont {A.}~\bibnamefont {Kreyssig}}, \bibinfo {author}
  {\bibfnamefont {M.}~\bibnamefont {Minola}}, \bibinfo {author} {\bibfnamefont
  {G.}~\bibnamefont {Dellea}}, \bibinfo {author} {\bibfnamefont
  {E.}~\bibnamefont {Weschke}}, \bibinfo {author} {\bibfnamefont {M.~J.}\
  \bibnamefont {Veit}}, \bibinfo {author} {\bibfnamefont {M.}~\bibnamefont
  {Ramazanoglu}}, \bibinfo {author} {\bibfnamefont {A.~I.}\ \bibnamefont
  {Goldman}}, \bibinfo {author} {\bibfnamefont {T.}~\bibnamefont {Schmitt}},
  \bibinfo {author} {\bibfnamefont {G.}~\bibnamefont {Ghiringhelli}}, \bibinfo
  {author} {\bibfnamefont {N.}~\bibnamefont {Bari\v{s}i\'c}}, \bibinfo {author}
  {\bibfnamefont {M.~K.}\ \bibnamefont {Chan}}, \bibinfo {author}
  {\bibfnamefont {C.~J.}\ \bibnamefont {Dorow}}, \bibinfo {author}
  {\bibfnamefont {G.}~\bibnamefont {Yu}}, \bibinfo {author} {\bibfnamefont
  {X.}~\bibnamefont {Zhao}}, \bibinfo {author} {\bibfnamefont {B.}~\bibnamefont
  {Keimer}}, \ and\ \bibinfo {author} {\bibfnamefont {M.}~\bibnamefont
  {Greven}},\ }\href@noop {} {\bibfield  {journal} {\bibinfo  {journal} {Nat.
  Commun.}\ }\textbf {\bibinfo {volume} {5}},\ \bibinfo {pages} {5875}
  (\bibinfo {year} {2014})}\BibitemShut {NoStop}%
\bibitem [{\citenamefont {Comin}\ \emph {et~al.}(2015)\citenamefont {Comin},
  \citenamefont {Sutarto}, \citenamefont {da~Silva~Neto}, \citenamefont
  {Chauviere}, \citenamefont {Liang}, \citenamefont {Hardy}, \citenamefont
  {Bonn}, \citenamefont {He}, \citenamefont {Sawatzky},\ and\ \citenamefont
  {Damascelli}}]{com2}%
  \BibitemOpen
  \bibfield  {author} {\bibinfo {author} {\bibfnamefont {R.}~\bibnamefont
  {Comin}}, \bibinfo {author} {\bibfnamefont {R.}~\bibnamefont {Sutarto}},
  \bibinfo {author} {\bibfnamefont {E.~H.}\ \bibnamefont {da~Silva~Neto}},
  \bibinfo {author} {\bibfnamefont {L.}~\bibnamefont {Chauviere}}, \bibinfo
  {author} {\bibfnamefont {R.}~\bibnamefont {Liang}}, \bibinfo {author}
  {\bibfnamefont {W.~N.}\ \bibnamefont {Hardy}}, \bibinfo {author}
  {\bibfnamefont {D.~A.}\ \bibnamefont {Bonn}}, \bibinfo {author}
  {\bibfnamefont {F.}~\bibnamefont {He}}, \bibinfo {author} {\bibfnamefont
  {G.~A.}\ \bibnamefont {Sawatzky}}, \ and\ \bibinfo {author} {\bibfnamefont
  {A.}~\bibnamefont {Damascelli}},\ }\href@noop {} {\bibfield  {journal}
  {\bibinfo  {journal} {Science}\ }\textbf {\bibinfo {volume} {347}},\ \bibinfo
  {pages} {1335} (\bibinfo {year} {2015})}\BibitemShut {NoStop}%
\bibitem [{\citenamefont {Hayward}\ \emph {et~al.}(2014)\citenamefont
  {Hayward}, \citenamefont {Hawthorn}, \citenamefont {Malko},\ and\
  \citenamefont {Sachdev}}]{hay1}%
  \BibitemOpen
  \bibfield  {author} {\bibinfo {author} {\bibfnamefont {L.~E.}\ \bibnamefont
  {Hayward}}, \bibinfo {author} {\bibfnamefont {D.~G.}\ \bibnamefont
  {Hawthorn}}, \bibinfo {author} {\bibfnamefont {R.~G.}\ \bibnamefont {Malko}},
  \ and\ \bibinfo {author} {\bibfnamefont {S.}~\bibnamefont {Sachdev}},\
  }\href@noop {} {\bibfield  {journal} {\bibinfo  {journal} {Science}\ }\textbf
  {\bibinfo {volume} {343}},\ \bibinfo {pages} {1336} (\bibinfo {year}
  {2014})}\BibitemShut {NoStop}%
\bibitem [{\citenamefont {Fujita}\ \emph {et~al.}(2014)\citenamefont {Fujita},
  \citenamefont {Kim}, \citenamefont {Lee}, \citenamefont {Hamidian},
  \citenamefont {Firmo}, \citenamefont {Mukhopadhyay}, \citenamefont {Eisaki},
  \citenamefont {Lawler}, \citenamefont {Kim},\ and\ \citenamefont
  {Davis}}]{fuj1}%
  \BibitemOpen
  \bibfield  {author} {\bibinfo {author} {\bibfnamefont {K.}~\bibnamefont
  {Fujita}}, \bibinfo {author} {\bibfnamefont {C.-K.}\ \bibnamefont {Kim}},
  \bibinfo {author} {\bibfnamefont {J.}~\bibnamefont {Lee}}, \bibinfo {author}
  {\bibfnamefont {M.~H.}\ \bibnamefont {Hamidian}}, \bibinfo {author}
  {\bibfnamefont {I.~A.}\ \bibnamefont {Firmo}}, \bibinfo {author}
  {\bibfnamefont {S.}~\bibnamefont {Mukhopadhyay}}, \bibinfo {author}
  {\bibfnamefont {H.}~\bibnamefont {Eisaki}}, \bibinfo {author} {\bibfnamefont
  {M.~J.}\ \bibnamefont {Lawler}}, \bibinfo {author} {\bibfnamefont {E.-A.}\
  \bibnamefont {Kim}}, \ and\ \bibinfo {author} {\bibfnamefont {J.~C.}\
  \bibnamefont {Davis}},\ }\href@noop {} {\bibfield  {journal} {\bibinfo
  {journal} {Science}\ }\textbf {\bibinfo {volume} {344}},\ \bibinfo {pages}
  {612} (\bibinfo {year} {2014})}\BibitemShut {NoStop}%
\bibitem [{\citenamefont {H\"ucker}\ \emph {et~al.}(2011)\citenamefont
  {H\"ucker}, \citenamefont {Zimmermann}, \citenamefont {Gu}, \citenamefont
  {Xu}, \citenamefont {Wen}, \citenamefont {Xu}, \citenamefont {Kang},
  \citenamefont {Zheludev},\ and\ \citenamefont {Tranquada}}]{huc1}%
  \BibitemOpen
  \bibfield  {author} {\bibinfo {author} {\bibfnamefont {M.}~\bibnamefont
  {H\"ucker}}, \bibinfo {author} {\bibfnamefont {M.~v.}\ \bibnamefont
  {Zimmermann}}, \bibinfo {author} {\bibfnamefont {G.~D.}\ \bibnamefont {Gu}},
  \bibinfo {author} {\bibfnamefont {Z.~J.}\ \bibnamefont {Xu}}, \bibinfo
  {author} {\bibfnamefont {J.~S.}\ \bibnamefont {Wen}}, \bibinfo {author}
  {\bibfnamefont {G.}~\bibnamefont {Xu}}, \bibinfo {author} {\bibfnamefont
  {H.~J.}\ \bibnamefont {Kang}}, \bibinfo {author} {\bibfnamefont
  {A.}~\bibnamefont {Zheludev}}, \ and\ \bibinfo {author} {\bibfnamefont
  {J.~M.}\ \bibnamefont {Tranquada}},\ }\href@noop {} {\bibfield  {journal}
  {\bibinfo  {journal} {Phys. Rev. B}\ }\textbf {\bibinfo {volume} {83}},\
  \bibinfo {pages} {104506} (\bibinfo {year} {2011})}\BibitemShut {NoStop}%
\bibitem [{\citenamefont {Fink}\ \emph {et~al.}(2011)\citenamefont {Fink},
  \citenamefont {Soltwisch}, \citenamefont {Geck}, \citenamefont {Schierle},
  \citenamefont {Weschke},\ and\ \citenamefont {B\"uchner}}]{fin1}%
  \BibitemOpen
  \bibfield  {author} {\bibinfo {author} {\bibfnamefont {J.}~\bibnamefont
  {Fink}}, \bibinfo {author} {\bibfnamefont {V.}~\bibnamefont {Soltwisch}},
  \bibinfo {author} {\bibfnamefont {J.}~\bibnamefont {Geck}}, \bibinfo {author}
  {\bibfnamefont {E.}~\bibnamefont {Schierle}}, \bibinfo {author}
  {\bibfnamefont {E.}~\bibnamefont {Weschke}}, \ and\ \bibinfo {author}
  {\bibfnamefont {B.}~\bibnamefont {B\"uchner}},\ }\href@noop {} {\bibfield
  {journal} {\bibinfo  {journal} {Phys. Rev. B}\ }\textbf {\bibinfo {volume}
  {83}},\ \bibinfo {pages} {092503} (\bibinfo {year} {2011})}\BibitemShut
  {NoStop}%
\bibitem [{\citenamefont {Crawford}\ \emph {et~al.}(1991)\citenamefont
  {Crawford}, \citenamefont {Harlow}, \citenamefont {McCarron}, \citenamefont
  {Farneth}, \citenamefont {Axe}, \citenamefont {Chou},\ and\ \citenamefont
  {Huang}}]{cra1}%
  \BibitemOpen
  \bibfield  {author} {\bibinfo {author} {\bibfnamefont {M.~K.}\ \bibnamefont
  {Crawford}}, \bibinfo {author} {\bibfnamefont {R.~L.}\ \bibnamefont
  {Harlow}}, \bibinfo {author} {\bibfnamefont {E.~M.}\ \bibnamefont
  {McCarron}}, \bibinfo {author} {\bibfnamefont {W.~E.}\ \bibnamefont
  {Farneth}}, \bibinfo {author} {\bibfnamefont {J.~D.}\ \bibnamefont {Axe}},
  \bibinfo {author} {\bibfnamefont {H.}~\bibnamefont {Chou}}, \ and\ \bibinfo
  {author} {\bibfnamefont {Q.}~\bibnamefont {Huang}},\ }\href@noop {}
  {\bibfield  {journal} {\bibinfo  {journal} {Phys. Rev. B}\ }\textbf {\bibinfo
  {volume} {44}},\ \bibinfo {pages} {7749} (\bibinfo {year}
  {1991})}\BibitemShut {NoStop}%
\bibitem [{\citenamefont {Suzuki}\ and\ \citenamefont {Fujita}(1989)}]{suz1}%
  \BibitemOpen
  \bibfield  {author} {\bibinfo {author} {\bibfnamefont {T.}~\bibnamefont
  {Suzuki}}\ and\ \bibinfo {author} {\bibfnamefont {T.}~\bibnamefont
  {Fujita}},\ }\href@noop {} {\bibfield  {journal} {\bibinfo  {journal}
  {Physica C}\ }\textbf {\bibinfo {volume} {159}},\ \bibinfo {pages} {111}
  (\bibinfo {year} {1989})}\BibitemShut {NoStop}%
\bibitem [{\citenamefont {Moodenbaugh}\ \emph {et~al.}(1988)\citenamefont
  {Moodenbaugh}, \citenamefont {Xu}, \citenamefont {Suenaga}, \citenamefont
  {Folkerts},\ and\ \citenamefont {Shelton}}]{moo1}%
  \BibitemOpen
  \bibfield  {author} {\bibinfo {author} {\bibfnamefont {A.~R.}\ \bibnamefont
  {Moodenbaugh}}, \bibinfo {author} {\bibfnamefont {Y.}~\bibnamefont {Xu}},
  \bibinfo {author} {\bibfnamefont {M.}~\bibnamefont {Suenaga}}, \bibinfo
  {author} {\bibfnamefont {T.~J.}\ \bibnamefont {Folkerts}}, \ and\ \bibinfo
  {author} {\bibfnamefont {R.~N.}\ \bibnamefont {Shelton}},\ }\href@noop {}
  {\bibfield  {journal} {\bibinfo  {journal} {Phys. Rev. B}\ }\textbf {\bibinfo
  {volume} {38}},\ \bibinfo {pages} {4596} (\bibinfo {year}
  {1988})}\BibitemShut {NoStop}%
\bibitem [{\citenamefont {H\"ucker}\ \emph {et~al.}(2010)\citenamefont
  {H\"ucker}, \citenamefont {Zimmermann}, \citenamefont {Debessai},
  \citenamefont {Schilling}, \citenamefont {Tranquada},\ and\ \citenamefont
  {D.}}]{huc2}%
  \BibitemOpen
  \bibfield  {author} {\bibinfo {author} {\bibfnamefont {M.}~\bibnamefont
  {H\"ucker}}, \bibinfo {author} {\bibfnamefont {M.~v.}\ \bibnamefont
  {Zimmermann}}, \bibinfo {author} {\bibfnamefont {M.}~\bibnamefont
  {Debessai}}, \bibinfo {author} {\bibfnamefont {J.~S.}\ \bibnamefont
  {Schilling}}, \bibinfo {author} {\bibfnamefont {J.~M.}\ \bibnamefont
  {Tranquada}}, \ and\ \bibinfo {author} {\bibfnamefont {G.~G.}\ \bibnamefont
  {D.}},\ }\href@noop {} {\bibfield  {journal} {\bibinfo  {journal} {Phys. Rev.
  Lett}\ }\textbf {\bibinfo {volume} {104}},\ \bibinfo {pages} {057004}
  (\bibinfo {year} {2010})}\BibitemShut {NoStop}%
\bibitem [{\citenamefont {Thampy}\ \emph {et~al.}(2014)\citenamefont {Thampy},
  \citenamefont {Dean}, \citenamefont {Christensen}, \citenamefont {Steinke},
  \citenamefont {Islam}, \citenamefont {Oda}, \citenamefont {Ido},
  \citenamefont {Momono}, \citenamefont {Wilkins},\ and\ \citenamefont
  {Hill}}]{tha1}%
  \BibitemOpen
  \bibfield  {author} {\bibinfo {author} {\bibfnamefont {V.}~\bibnamefont
  {Thampy}}, \bibinfo {author} {\bibfnamefont {M.~P.~M.}\ \bibnamefont {Dean}},
  \bibinfo {author} {\bibfnamefont {N.~B.}\ \bibnamefont {Christensen}},
  \bibinfo {author} {\bibfnamefont {L.}~\bibnamefont {Steinke}}, \bibinfo
  {author} {\bibfnamefont {Z.}~\bibnamefont {Islam}}, \bibinfo {author}
  {\bibfnamefont {M.}~\bibnamefont {Oda}}, \bibinfo {author} {\bibfnamefont
  {M.}~\bibnamefont {Ido}}, \bibinfo {author} {\bibfnamefont {N.}~\bibnamefont
  {Momono}}, \bibinfo {author} {\bibfnamefont {S.~B.}\ \bibnamefont {Wilkins}},
  \ and\ \bibinfo {author} {\bibfnamefont {J.~P.}\ \bibnamefont {Hill}},\
  }\href@noop {} {\bibfield  {journal} {\bibinfo  {journal} {Phys. Rev. B}\
  }\textbf {\bibinfo {volume} {90}},\ \bibinfo {pages} {100510(R)} (\bibinfo
  {year} {2014})}\BibitemShut {NoStop}%
\bibitem [{\citenamefont {F\"orst}\ \emph
  {et~al.}(2014{\natexlab{a}})\citenamefont {F\"orst}, \citenamefont {Tobey},
  \citenamefont {Bromberger}, \citenamefont {Wilkins}, \citenamefont {Khanna},
  \citenamefont {Caviglia}, \citenamefont {Chuang}, \citenamefont {Lee},
  \citenamefont {Schlotter}, \citenamefont {Turner}, \citenamefont {Minitti},
  \citenamefont {Krupin}, \citenamefont {Xu}, \citenamefont {Wen},
  \citenamefont {Gu}, \citenamefont {Dhesi}, \citenamefont {Cavalleri},\ and\
  \citenamefont {Hill}}]{for1}%
  \BibitemOpen
  \bibfield  {author} {\bibinfo {author} {\bibfnamefont {M.}~\bibnamefont
  {F\"orst}}, \bibinfo {author} {\bibfnamefont {R.~I.}\ \bibnamefont {Tobey}},
  \bibinfo {author} {\bibfnamefont {H.}~\bibnamefont {Bromberger}}, \bibinfo
  {author} {\bibfnamefont {S.~B.}\ \bibnamefont {Wilkins}}, \bibinfo {author}
  {\bibfnamefont {V.}~\bibnamefont {Khanna}}, \bibinfo {author} {\bibfnamefont
  {A.~D.}\ \bibnamefont {Caviglia}}, \bibinfo {author} {\bibfnamefont {Y.-D.}\
  \bibnamefont {Chuang}}, \bibinfo {author} {\bibfnamefont {W.~S.}\
  \bibnamefont {Lee}}, \bibinfo {author} {\bibfnamefont {W.~F.}\ \bibnamefont
  {Schlotter}}, \bibinfo {author} {\bibfnamefont {J.~J.}\ \bibnamefont
  {Turner}}, \bibinfo {author} {\bibfnamefont {M.~P.}\ \bibnamefont {Minitti}},
  \bibinfo {author} {\bibfnamefont {O.}~\bibnamefont {Krupin}}, \bibinfo
  {author} {\bibfnamefont {Z.~J.}\ \bibnamefont {Xu}}, \bibinfo {author}
  {\bibfnamefont {J.~S.}\ \bibnamefont {Wen}}, \bibinfo {author} {\bibfnamefont
  {G.~D.}\ \bibnamefont {Gu}}, \bibinfo {author} {\bibfnamefont {S.~S.}\
  \bibnamefont {Dhesi}}, \bibinfo {author} {\bibfnamefont {A.}~\bibnamefont
  {Cavalleri}}, \ and\ \bibinfo {author} {\bibfnamefont {J.~P.}\ \bibnamefont
  {Hill}},\ }\href@noop {} {\bibfield  {journal} {\bibinfo  {journal} {Phys.
  Rev. Lett.}\ }\textbf {\bibinfo {volume} {112}},\ \bibinfo {pages} {157002}
  (\bibinfo {year} {2014}{\natexlab{a}})}\BibitemShut {NoStop}%
\bibitem [{\citenamefont {Li}\ \emph {et~al.}(2007)\citenamefont {Li},
  \citenamefont {H\"ucker}, \citenamefont {Gu}, \citenamefont {Tsvelik},\ and\
  \citenamefont {Tranquada}}]{li1}%
  \BibitemOpen
  \bibfield  {author} {\bibinfo {author} {\bibfnamefont {Q.}~\bibnamefont
  {Li}}, \bibinfo {author} {\bibfnamefont {M.}~\bibnamefont {H\"ucker}},
  \bibinfo {author} {\bibfnamefont {G.~D.}\ \bibnamefont {Gu}}, \bibinfo
  {author} {\bibfnamefont {A.~M.}\ \bibnamefont {Tsvelik}}, \ and\ \bibinfo
  {author} {\bibfnamefont {J.~M.}\ \bibnamefont {Tranquada}},\ }\href@noop {}
  {\bibfield  {journal} {\bibinfo  {journal} {Phys. Rev. Lett.}\ }\textbf
  {\bibinfo {volume} {99}},\ \bibinfo {pages} {067001} (\bibinfo {year}
  {2007})}\BibitemShut {NoStop}%
\bibitem [{\citenamefont {Berg}\ \emph {et~al.}(2007)\citenamefont {Berg},
  \citenamefont {Fradkin}, \citenamefont {Kim}, \citenamefont {Kivelson},
  \citenamefont {Oganesyan}, \citenamefont {Tranquada},\ and\ \citenamefont
  {Zhang}}]{ber1}%
  \BibitemOpen
  \bibfield  {author} {\bibinfo {author} {\bibfnamefont {E.}~\bibnamefont
  {Berg}}, \bibinfo {author} {\bibfnamefont {E.}~\bibnamefont {Fradkin}},
  \bibinfo {author} {\bibfnamefont {E.-A.}\ \bibnamefont {Kim}}, \bibinfo
  {author} {\bibfnamefont {S.~A.}\ \bibnamefont {Kivelson}}, \bibinfo {author}
  {\bibfnamefont {V.}~\bibnamefont {Oganesyan}}, \bibinfo {author}
  {\bibfnamefont {J.~M.}\ \bibnamefont {Tranquada}}, \ and\ \bibinfo {author}
  {\bibfnamefont {S.~C.}\ \bibnamefont {Zhang}},\ }\href@noop {} {\bibfield
  {journal} {\bibinfo  {journal} {Phys. Rev. Lett.}\ }\textbf {\bibinfo
  {volume} {99}},\ \bibinfo {pages} {127003} (\bibinfo {year}
  {2007})}\BibitemShut {NoStop}%
\bibitem [{\citenamefont {Berg}\ \emph {et~al.}(2009)\citenamefont {Berg},
  \citenamefont {Fradkin},\ and\ \citenamefont {Kivelson}}]{he1}%
  \BibitemOpen
  \bibfield  {author} {\bibinfo {author} {\bibfnamefont {E.}~\bibnamefont
  {Berg}}, \bibinfo {author} {\bibfnamefont {E.}~\bibnamefont {Fradkin}}, \
  and\ \bibinfo {author} {\bibfnamefont {S.~A.}\ \bibnamefont {Kivelson}},\
  }\href@noop {} {\bibfield  {journal} {\bibinfo  {journal} {Nature Phys.}\
  }\textbf {\bibinfo {volume} {5}},\ \bibinfo {pages} {830} (\bibinfo {year}
  {2009})}\BibitemShut {NoStop}%
\bibitem [{\citenamefont {Abbamonte}\ \emph {et~al.}(2005)\citenamefont
  {Abbamonte}, \citenamefont {Rusydi}, \citenamefont {Smadici}, \citenamefont
  {Gu}, \citenamefont {Sawatzky},\ and\ \citenamefont {Feng}}]{abb1}%
  \BibitemOpen
  \bibfield  {author} {\bibinfo {author} {\bibfnamefont {P.}~\bibnamefont
  {Abbamonte}}, \bibinfo {author} {\bibfnamefont {A.}~\bibnamefont {Rusydi}},
  \bibinfo {author} {\bibfnamefont {S.}~\bibnamefont {Smadici}}, \bibinfo
  {author} {\bibfnamefont {G.~D.}\ \bibnamefont {Gu}}, \bibinfo {author}
  {\bibfnamefont {G.~A.}\ \bibnamefont {Sawatzky}}, \ and\ \bibinfo {author}
  {\bibfnamefont {D.~L.}\ \bibnamefont {Feng}},\ }\href@noop {} {\bibfield
  {journal} {\bibinfo  {journal} {Nature Phys.}\ }\textbf {\bibinfo {volume}
  {1}},\ \bibinfo {pages} {155} (\bibinfo {year} {2005})}\BibitemShut {NoStop}%
\bibitem [{\citenamefont {Homes}\ \emph {et~al.}(2012)\citenamefont {Homes},
  \citenamefont {H\"ucker}, \citenamefont {Li}, \citenamefont {Xu},
  \citenamefont {Wen}, \citenamefont {Gu},\ and\ \citenamefont
  {Tranquada}}]{hom1}%
  \BibitemOpen
  \bibfield  {author} {\bibinfo {author} {\bibfnamefont {C.~C.}\ \bibnamefont
  {Homes}}, \bibinfo {author} {\bibfnamefont {M.}~\bibnamefont {H\"ucker}},
  \bibinfo {author} {\bibfnamefont {Q.}~\bibnamefont {Li}}, \bibinfo {author}
  {\bibfnamefont {Z.~J.}\ \bibnamefont {Xu}}, \bibinfo {author} {\bibfnamefont
  {J.~S.}\ \bibnamefont {Wen}}, \bibinfo {author} {\bibfnamefont {G.~D.}\
  \bibnamefont {Gu}}, \ and\ \bibinfo {author} {\bibfnamefont {J.~M.}\
  \bibnamefont {Tranquada}},\ }\href@noop {} {\bibfield  {journal} {\bibinfo
  {journal} {Phys. Rev. B}\ }\textbf {\bibinfo {volume} {85}},\ \bibinfo
  {pages} {134510} (\bibinfo {year} {2012})}\BibitemShut {NoStop}%
\bibitem [{\citenamefont {Kao}\ \emph {et~al.}(1989)\citenamefont {Kao},
  \citenamefont {Kwok},\ and\ \citenamefont {Shaw}}]{han1}%
  \BibitemOpen
  \bibfield  {author} {\bibinfo {author} {\bibfnamefont {Y.-H.}\ \bibnamefont
  {Kao}}, \bibinfo {author} {\bibfnamefont {H.}~\bibnamefont {Kwok}}, \ and\
  \bibinfo {author} {\bibfnamefont {D.~T.}\ \bibnamefont {Shaw}},\ }\href@noop
  {} {\emph {\bibinfo {title} {Superconductivity and Applications}}},\ edited
  by\ \bibinfo {editor} {\bibfnamefont {Y.-H.}\ \bibnamefont {Kao}}, \bibinfo
  {editor} {\bibfnamefont {H.}~\bibnamefont {Kwok}}, \ and\ \bibinfo {editor}
  {\bibfnamefont {D.~T.}\ \bibnamefont {Shaw}}\ (\bibinfo  {publisher}
  {Springer},\ \bibinfo {year} {1989})\ p.\ \bibinfo {pages} {281}\BibitemShut
  {NoStop}%
\bibitem [{\citenamefont {Fink}\ \emph {et~al.}(2009)\citenamefont {Fink},
  \citenamefont {Schierle}, \citenamefont {Weschke}, \citenamefont {Geck},
  \citenamefont {Hawthorn}, \citenamefont {Soltwisch}, \citenamefont {Wadati},
  \citenamefont {Wu}, \citenamefont {D\"urr}, \citenamefont {Wizent},
  \citenamefont {B\"uchner},\ and\ \citenamefont {Swatzky}}]{fin2}%
  \BibitemOpen
  \bibfield  {author} {\bibinfo {author} {\bibfnamefont {J.}~\bibnamefont
  {Fink}}, \bibinfo {author} {\bibfnamefont {E.}~\bibnamefont {Schierle}},
  \bibinfo {author} {\bibfnamefont {E.}~\bibnamefont {Weschke}}, \bibinfo
  {author} {\bibfnamefont {J.}~\bibnamefont {Geck}}, \bibinfo {author}
  {\bibfnamefont {D.}~\bibnamefont {Hawthorn}}, \bibinfo {author}
  {\bibfnamefont {V.}~\bibnamefont {Soltwisch}}, \bibinfo {author}
  {\bibfnamefont {H.}~\bibnamefont {Wadati}}, \bibinfo {author} {\bibfnamefont
  {H.-H.}\ \bibnamefont {Wu}}, \bibinfo {author} {\bibfnamefont {H.~A.}\
  \bibnamefont {D\"urr}}, \bibinfo {author} {\bibfnamefont {N.}~\bibnamefont
  {Wizent}}, \bibinfo {author} {\bibfnamefont {B.}~\bibnamefont {B\"uchner}}, \
  and\ \bibinfo {author} {\bibfnamefont {G.~A.}\ \bibnamefont {Swatzky}},\
  }\href@noop {} {\bibfield  {journal} {\bibinfo  {journal} {Phys. Rev. B}\
  }\textbf {\bibinfo {volume} {79}},\ \bibinfo {pages} {100502} (\bibinfo
  {year} {2009})}\BibitemShut {NoStop}%
\bibitem [{\citenamefont {Fausti}\ \emph {et~al.}(2011)\citenamefont {Fausti},
  \citenamefont {Tobey}, \citenamefont {Dean}, \citenamefont {Kaiser},
  \citenamefont {Dienst}, \citenamefont {Hoffmann}, \citenamefont {Pyon},
  \citenamefont {Takayama}, \citenamefont {Takagi},\ and\ \citenamefont
  {Cavalleri}}]{fau1}%
  \BibitemOpen
  \bibfield  {author} {\bibinfo {author} {\bibfnamefont {D.}~\bibnamefont
  {Fausti}}, \bibinfo {author} {\bibfnamefont {R.~I.}\ \bibnamefont {Tobey}},
  \bibinfo {author} {\bibfnamefont {N.}~\bibnamefont {Dean}}, \bibinfo {author}
  {\bibfnamefont {S.}~\bibnamefont {Kaiser}}, \bibinfo {author} {\bibfnamefont
  {A.}~\bibnamefont {Dienst}}, \bibinfo {author} {\bibfnamefont {M.~C.}\
  \bibnamefont {Hoffmann}}, \bibinfo {author} {\bibfnamefont {S.}~\bibnamefont
  {Pyon}}, \bibinfo {author} {\bibfnamefont {T.}~\bibnamefont {Takayama}},
  \bibinfo {author} {\bibfnamefont {H.}~\bibnamefont {Takagi}}, \ and\ \bibinfo
  {author} {\bibfnamefont {A.}~\bibnamefont {Cavalleri}},\ }\href@noop {}
  {\bibfield  {journal} {\bibinfo  {journal} {Science}\ }\textbf {\bibinfo
  {volume} {331}},\ \bibinfo {pages} {189} (\bibinfo {year}
  {2011})}\BibitemShut {NoStop}%
\bibitem [{\citenamefont {Hu}\ \emph {et~al.}(2014)\citenamefont {Hu},
  \citenamefont {Kaiser}, \citenamefont {Nicoletti}, \citenamefont {Hunt},
  \citenamefont {Gierz}, \citenamefont {Hoffmann}, \citenamefont {Le~Tacon},
  \citenamefont {Loew}, \citenamefont {Keimer},\ and\ \citenamefont
  {Cavalleri}}]{hu1}%
  \BibitemOpen
  \bibfield  {author} {\bibinfo {author} {\bibfnamefont {W.}~\bibnamefont
  {Hu}}, \bibinfo {author} {\bibfnamefont {S.}~\bibnamefont {Kaiser}}, \bibinfo
  {author} {\bibfnamefont {D.}~\bibnamefont {Nicoletti}}, \bibinfo {author}
  {\bibfnamefont {C.~R.}\ \bibnamefont {Hunt}}, \bibinfo {author}
  {\bibfnamefont {I.}~\bibnamefont {Gierz}}, \bibinfo {author} {\bibfnamefont
  {M.~C.}\ \bibnamefont {Hoffmann}}, \bibinfo {author} {\bibfnamefont
  {M.}~\bibnamefont {Le~Tacon}}, \bibinfo {author} {\bibfnamefont
  {T.}~\bibnamefont {Loew}}, \bibinfo {author} {\bibfnamefont {B.}~\bibnamefont
  {Keimer}}, \ and\ \bibinfo {author} {\bibfnamefont {A.}~\bibnamefont
  {Cavalleri}},\ }\href@noop {} {\bibfield  {journal} {\bibinfo  {journal}
  {Nat. Mater.}\ }\textbf {\bibinfo {volume} {13}},\ \bibinfo {pages} {705}
  (\bibinfo {year} {2014})}\BibitemShut {NoStop}%
\bibitem [{\citenamefont {F\"orst}\ \emph
  {et~al.}(2014{\natexlab{b}})\citenamefont {F\"orst}, \citenamefont {Frano},
  \citenamefont {Kaiser}, \citenamefont {Mankowsky}, \citenamefont {Hunt},
  \citenamefont {Turner}, \citenamefont {Dakovski}, \citenamefont {Minitti},
  \citenamefont {Robinson}, \citenamefont {Loew}, \citenamefont {Le~Tacon},
  \citenamefont {Keimer}, \citenamefont {Hill}, \citenamefont {Cavalleri},\
  and\ \citenamefont {Dhesi}}]{for2}%
  \BibitemOpen
  \bibfield  {author} {\bibinfo {author} {\bibfnamefont {M.}~\bibnamefont
  {F\"orst}}, \bibinfo {author} {\bibfnamefont {A.}~\bibnamefont {Frano}},
  \bibinfo {author} {\bibfnamefont {S.}~\bibnamefont {Kaiser}}, \bibinfo
  {author} {\bibfnamefont {R.}~\bibnamefont {Mankowsky}}, \bibinfo {author}
  {\bibfnamefont {C.~R.}\ \bibnamefont {Hunt}}, \bibinfo {author}
  {\bibfnamefont {J.~J.}\ \bibnamefont {Turner}}, \bibinfo {author}
  {\bibfnamefont {G.~L.}\ \bibnamefont {Dakovski}}, \bibinfo {author}
  {\bibfnamefont {M.~P.}\ \bibnamefont {Minitti}}, \bibinfo {author}
  {\bibfnamefont {J.}~\bibnamefont {Robinson}}, \bibinfo {author}
  {\bibfnamefont {T.}~\bibnamefont {Loew}}, \bibinfo {author} {\bibfnamefont
  {M.}~\bibnamefont {Le~Tacon}}, \bibinfo {author} {\bibfnamefont
  {B.}~\bibnamefont {Keimer}}, \bibinfo {author} {\bibfnamefont {J.~P.}\
  \bibnamefont {Hill}}, \bibinfo {author} {\bibfnamefont {A.}~\bibnamefont
  {Cavalleri}}, \ and\ \bibinfo {author} {\bibfnamefont {S.~S.}\ \bibnamefont
  {Dhesi}},\ }\href@noop {} {\bibfield  {journal} {\bibinfo  {journal} {Phys.
  Rev. B}\ }\textbf {\bibinfo {volume} {90}},\ \bibinfo {pages} {184514}
  (\bibinfo {year} {2014}{\natexlab{b}})}\BibitemShut {NoStop}%
\bibitem [{equ()}]{equJPR}%
  \BibitemOpen
  \href@noop {} {}\bibinfo {note} {The THz-frequency reflectivity ratios
  measured above and below T$_c$ were combined with the broadband (up to 10,000
  cm$^{-1}$) reflectivity. By applying Kramers-Kronig transformations, a full
  set of equilibrium optical properties at T=5K could be
  determined}\BibitemShut {NoStop}%
\bibitem [{\citenamefont {Nicoletti}\ \emph {et~al.}(2014)\citenamefont
  {Nicoletti}, \citenamefont {Casandruc}, \citenamefont {Laplace},
  \citenamefont {Khanna}, \citenamefont {Hunt}, \citenamefont {Kaiser},
  \citenamefont {Dhesi}, \citenamefont {Gu}, \citenamefont {Hill},\ and\
  \citenamefont {Cavalleri}}]{nic1}%
  \BibitemOpen
  \bibfield  {author} {\bibinfo {author} {\bibfnamefont {D.}~\bibnamefont
  {Nicoletti}}, \bibinfo {author} {\bibfnamefont {E.}~\bibnamefont
  {Casandruc}}, \bibinfo {author} {\bibfnamefont {Y.}~\bibnamefont {Laplace}},
  \bibinfo {author} {\bibfnamefont {V.}~\bibnamefont {Khanna}}, \bibinfo
  {author} {\bibfnamefont {C.~R.}\ \bibnamefont {Hunt}}, \bibinfo {author}
  {\bibfnamefont {S.}~\bibnamefont {Kaiser}}, \bibinfo {author} {\bibfnamefont
  {S.~S.}\ \bibnamefont {Dhesi}}, \bibinfo {author} {\bibfnamefont {G.~D.}\
  \bibnamefont {Gu}}, \bibinfo {author} {\bibfnamefont {J.~P.}\ \bibnamefont
  {Hill}}, \ and\ \bibinfo {author} {\bibfnamefont {A.}~\bibnamefont
  {Cavalleri}},\ }\href@noop {} {\bibfield  {journal} {\bibinfo  {journal}
  {Phys. Rev. B}\ }\textbf {\bibinfo {volume} {90}},\ \bibinfo {pages} {100503}
  (\bibinfo {year} {2014})}\BibitemShut {NoStop}%
\bibitem [{mis()}]{mismatch1}%
  \BibitemOpen
  \href@noop {} {}\bibinfo {note} {The penetration depth mismatch was taken
  into account by modelling the response of the system as that of a
  homogeneously photo-excited thin layer on top of an unperturbed bulk which
  retains the equilibrium optical properties. By calculating the coupled
  Fresnel equations of such a multi-layer, the transient optical response of
  the photoexcited layer could be derived.}\BibitemShut {Stop}%
\bibitem [{\citenamefont {Casandruc}\ \emph {et~al.}(2015)\citenamefont
  {Casandruc}, \citenamefont {Nicoletti}, \citenamefont {Rajasekaran},
  \citenamefont {Laplace}, \citenamefont {Khanna}, \citenamefont {Gu},
  \citenamefont {Hill},\ and\ \citenamefont {Cavalleri}}]{cas1}%
  \BibitemOpen
  \bibfield  {author} {\bibinfo {author} {\bibfnamefont {E.}~\bibnamefont
  {Casandruc}}, \bibinfo {author} {\bibfnamefont {D.}~\bibnamefont
  {Nicoletti}}, \bibinfo {author} {\bibfnamefont {S.}~\bibnamefont
  {Rajasekaran}}, \bibinfo {author} {\bibfnamefont {Y.}~\bibnamefont
  {Laplace}}, \bibinfo {author} {\bibfnamefont {V.}~\bibnamefont {Khanna}},
  \bibinfo {author} {\bibfnamefont {G.~D.}\ \bibnamefont {Gu}}, \bibinfo
  {author} {\bibfnamefont {J.~P.}\ \bibnamefont {Hill}}, \ and\ \bibinfo
  {author} {\bibfnamefont {A.}~\bibnamefont {Cavalleri}},\ }\href@noop {}
  {\bibfield  {journal} {\bibinfo  {journal} {Phys. Rev. B}\ }\textbf {\bibinfo
  {volume} {91}},\ \bibinfo {pages} {174502} (\bibinfo {year}
  {2015})}\BibitemShut {NoStop}%
\bibitem [{\citenamefont {Kusar}\ \emph {et~al.}(2008)\citenamefont {Kusar},
  \citenamefont {Kabanov}, \citenamefont {Demsar}, \citenamefont {Mertelj},
  \citenamefont {Sugai},\ and\ \citenamefont {Mihailovic}}]{kus1}%
  \BibitemOpen
  \bibfield  {author} {\bibinfo {author} {\bibfnamefont {P.}~\bibnamefont
  {Kusar}}, \bibinfo {author} {\bibfnamefont {V.~V.}\ \bibnamefont {Kabanov}},
  \bibinfo {author} {\bibfnamefont {J.}~\bibnamefont {Demsar}}, \bibinfo
  {author} {\bibfnamefont {T.}~\bibnamefont {Mertelj}}, \bibinfo {author}
  {\bibfnamefont {S.}~\bibnamefont {Sugai}}, \ and\ \bibinfo {author}
  {\bibfnamefont {D.}~\bibnamefont {Mihailovic}},\ }\href@noop {} {\bibfield
  {journal} {\bibinfo  {journal} {Phys. Rev. Lett.}\ }\textbf {\bibinfo
  {volume} {101}},\ \bibinfo {pages} {227001} (\bibinfo {year}
  {2008})}\BibitemShut {NoStop}%
\bibitem [{\citenamefont {Smallwood}\ \emph {et~al.}(2012)\citenamefont
  {Smallwood}, \citenamefont {Hinton}, \citenamefont {Jozwiak}, \citenamefont
  {Zhang}, \citenamefont {Koralek}, \citenamefont {Eisaki}, \citenamefont
  {Lee}, \citenamefont {Orenstein},\ and\ \citenamefont {Lanzara}}]{sma1}%
  \BibitemOpen
  \bibfield  {author} {\bibinfo {author} {\bibfnamefont {C.~L.}\ \bibnamefont
  {Smallwood}}, \bibinfo {author} {\bibfnamefont {J.~P.}\ \bibnamefont
  {Hinton}}, \bibinfo {author} {\bibfnamefont {C.}~\bibnamefont {Jozwiak}},
  \bibinfo {author} {\bibfnamefont {W.}~\bibnamefont {Zhang}}, \bibinfo
  {author} {\bibfnamefont {J.~D.}\ \bibnamefont {Koralek}}, \bibinfo {author}
  {\bibfnamefont {H.}~\bibnamefont {Eisaki}}, \bibinfo {author} {\bibfnamefont
  {D.-H.}\ \bibnamefont {Lee}}, \bibinfo {author} {\bibfnamefont
  {J.}~\bibnamefont {Orenstein}}, \ and\ \bibinfo {author} {\bibfnamefont
  {A.}~\bibnamefont {Lanzara}},\ }\href@noop {} {\bibfield  {journal} {\bibinfo
   {journal} {Science}\ }\textbf {\bibinfo {volume} {336}},\ \bibinfo {pages}
  {1137} (\bibinfo {year} {2012})}\BibitemShut {NoStop}%
\bibitem [{\citenamefont {Cort\'es}\ \emph {et~al.}(2011)\citenamefont
  {Cort\'es}, \citenamefont {Rettig}, \citenamefont {Yoshida}, \citenamefont
  {Eisaki}, \citenamefont {Wolf},\ and\ \citenamefont {Bovensiepen}}]{cor1}%
  \BibitemOpen
  \bibfield  {author} {\bibinfo {author} {\bibfnamefont {R.}~\bibnamefont
  {Cort\'es}}, \bibinfo {author} {\bibfnamefont {L.}~\bibnamefont {Rettig}},
  \bibinfo {author} {\bibfnamefont {Y.}~\bibnamefont {Yoshida}}, \bibinfo
  {author} {\bibfnamefont {H.}~\bibnamefont {Eisaki}}, \bibinfo {author}
  {\bibfnamefont {M.}~\bibnamefont {Wolf}}, \ and\ \bibinfo {author}
  {\bibfnamefont {U.}~\bibnamefont {Bovensiepen}},\ }\href@noop {} {\bibfield
  {journal} {\bibinfo  {journal} {Phys. Rev. Lett.}\ }\textbf {\bibinfo
  {volume} {107}},\ \bibinfo {pages} {097002} (\bibinfo {year}
  {2011})}\BibitemShut {NoStop}%
\bibitem [{\citenamefont {Magnuson}\ \emph {et~al.}(2015)\citenamefont
  {Magnuson}, \citenamefont {Schmitt}, \citenamefont {Strocov}, \citenamefont
  {Schlappa}, \citenamefont {Kalabukhov},\ and\ \citenamefont {Duda}}]{mag1}%
  \BibitemOpen
  \bibfield  {author} {\bibinfo {author} {\bibfnamefont {M.}~\bibnamefont
  {Magnuson}}, \bibinfo {author} {\bibfnamefont {T.}~\bibnamefont {Schmitt}},
  \bibinfo {author} {\bibfnamefont {V.~N.}\ \bibnamefont {Strocov}}, \bibinfo
  {author} {\bibfnamefont {J.}~\bibnamefont {Schlappa}}, \bibinfo {author}
  {\bibfnamefont {A.~S.}\ \bibnamefont {Kalabukhov}}, \ and\ \bibinfo {author}
  {\bibfnamefont {L.-C.}\ \bibnamefont {Duda}},\ }\href@noop {} {\bibfield
  {journal} {\bibinfo  {journal} {Sci. Rep.}\ }\textbf {\bibinfo {volume}
  {4}},\ \bibinfo {pages} {7017} (\bibinfo {year} {2015})}\BibitemShut
  {NoStop}%
\end{thebibliography}%

\end{document}